\documentclass[final]{elsarticle_ARXIV} % ELSEVIER

\usepackage{lineno,hyperref}
\usepackage{color}
\definecolor{green1}{RGB}{0.0,  0.0, 153}
\usepackage{geometry}
\begin{document}

	 \begin{frontmatter} % ELSEVIER
	
	\title{Reactive transport under stress: Permeability evolution in deformable porous media}

%	% &&& AGU AUTHORSHIP
%	\authors{R. Roded\affil{1},
%		X. Paredes\affil{2}\thanks{Current address, Faculty of Sciences, University of Lisbon, Portugal}, R. Holtzman\affil{2}}
%	
%	\affiliation{1}{Hydrology and Water Resources Program, The Hebrew University of Jerusalem, Jerusalem, Israel}
%	\affiliation{2}{Department of Soil and Water Sciences, The Hebrew University of Jerusalem, Rehovot, Israel}
%
%	\correspondingauthor{R. Holtzman}{holtzman.ran@mail.huji.ac.il}
%	% &&& AGU AUTHORSHIP

 % &&& ELSEVIER AUTHORSHIP 		
% %% Group authors per affiliation:
%	\author{Roi Roded, Xavier Paredes, Ran Holtzman\fnref{myfootnote}}
%	\address{Department of Soil and Water Sciences, The Hebrew University of Jerusalem, Rehovot 7610001, Israel}
%	\fntext[myfootnote]{Since 1880.}
	
	%% ELSEVIER
	\author[mymainaddress]{R. Roded}
	%\ead{roi.roded@mail.huji.ac.il}
	
	\author[mymainaddress,mysecondaryaddress]{X. Paredes}
	%\ead{xavierparedesmendez@gmail.com}
	
	\author[mymainaddress]{R. Holtzman\corref{mycorrespondingauthor}}
	\cortext[mycorrespondingauthor]{Corresponding author}
	\ead{holtzman.ran@mail.huji.ac.il}

	\address[mymainaddress]{
%	Department of Soil and Water Sciences, 
	The Hebrew University of Jerusalem, Rehovot 7610001, Israel}
	\address[mysecondaryaddress]{Faculty of Sciences, University of Lisbon, Portugal}
 % &&& ELSEVIER AUTHORSHIP 		
	
%	%%=== 	% AGU keypoints ===%%	
%	\begin{keypoints}
%		\item A pore-scale model captures the coupling of dissolution and weakening-induced mechanical compaction
%		\item Increasing stress leads to stronger inhibition of the permeability evolution due to weakening upstream and stress concentration downstream
%		\item At the wormholing regime (high $Da$), stress reduces transport heterogeneity, increasing wormhole competition
%	\end{keypoints}
%	%%=== 	% AGU keypoints ===%%	
%	% AGU
%	
	
	%%=== ABSTRACT ===%%	
	\begin{abstract}
		We study reactive transport in a stressed porous media, where dissolution of the solid matrix causes two simultaneous, competing effects: pore enlargement due to chemical deformation, and pore compaction due to mechanical weakening. We use a novel, mechanistic pore-scale model to simulate flooding of a sample under fixed confining stress. Our simulations show that increasing the stress inhibits the permeability enhancement, increasing the injected volume required to reach a certain permeability, in agreement with recent experiments. We explain this behavior by stress concentration downstream, in the less dissolved (hence stiffer) outlet region. As this region is also less conductive, even its small compaction has a strong bottleneck effect that curbs the permeability. 
		
Our results also elucidate that the impact of stress depends on the dissolution regime. Under wormholing conditions (slow injection, i.e. high Damkohler number, $Da$), the development of a sharp dissolution front and high porosity contrast accentuates the bottleneck effect. This reduces transport heterogeneity, promoting wormhole competition. Once the outlet starts eroding, the extreme focusing of transport and hence dissolution---characteristic of wormholing---becomes dominant, diminishing the bottleneck effect and hence the impact of stress at breakthrough. In contrast, at high flow rates (low $Da$), incomplete reaction upstream allows some of the reagent to traverse the sample, causing a more uniform dissolution. The continuous dissolution and its partial counteraction by compaction at the outlet provides a steady, gradual increase in the effect of stress. Consequently, the impact of stress is more pronounced at high $Da$ during early stages (low permeability), and at low $Da$ close to breakthrough. Our work promotes understanding of the interplay between dissolution and compaction and its effect on the hydromechanical property evolution, with important implications for processes ranging from diagenesis and weathering of rocks, to well stimulation and carbon geosequetration.
	\end{abstract}
	%%=== ABSTRACT ===%%

% %%%%% ARXIV
% % &&& ELSEVIER keyword 		
%\begin{keyword}
%Hydro-chemo-mechanical coupling \sep Mineral dissolution \sep Pore-scale simulations \sep Wormholing \sep Mechanical compaction \sep Permeability evolution
%\end{keyword}
%	 % &&& ELSEVIER keyword 		
%%%%%% ARXIV
	
	\end{frontmatter}% ELSEVIER
%%%%% ARXIV	\linenumbers

	% ==================================================
	% ==================================================
	% ==================================================
	\section{Introduction}\label{intro}
	% ==================================================
	% ==================================================
	% ==================================================
	
		%% ###### broad context--RT is complex, important ###### %%
	
	The transport of reactive fluids in porous media is a fundamental process in many earth systems, including weathering and diagenesis of rocks, contaminant transport, subsurface remediation, energy recovery, and CO$_2$ sequestration. 
	Reactive transport is a complex, nonlinear process: the transport and reactive properties and hence the rate and spatial distribution of fluid flow and reaction strongly depend on the microstructure, which in turn keeps evolving in time with the reaction~\citep{SteefelLasaga, Dentz2011,Noiriel2015}.

			%% ###### more specifically, different behaviors, permability ###### %%

	Much interest has been given to the dissolution of the solid matrix and the consequent evolution of the porosity and permeability.
	%, which control the transport. 
%	
%	In particular, there is an extensive body of literature on the evolution of the transport properties, pointing to significant variations in the behavior (different regimes) depending on the hydrochemical conditions. 
	%
Experiments and numerical simulations pointed out a rich range of behavior, or regimes, depending on the hydrochemical conditions. It has been shown that for reactions which are slow relative to the rate of fluid flow (the ratio being the Damkohler number, $Da$), ample reagent is provided throughout the sample, leading to relatively spatially-uniform dissolution, and steady enhancement of permeability. In contrast, at higher $Da$ (the so-called transport-limited regime), most of the reagent is consumed by preferential dissolution of the most conductive flow paths (``wormholes''). This further enhances their conductivity relative to other, less conductive regions, leading eventually to runaway permeability increase~\citep{Daccord_nature1987, Bekri_CES1995, Hoefner_AIChE1988, Szymczak_JGR2009}.

	%% ###### RT UNDER STRESS, in general ###### %%
	
	In many cases, for instance in the subsurface, the medium is under appreciable stress, such that dissolution (``chemical deformation'') also induces mechanical deformation. 
	Chemomechanical deformation and the coupled changes in the transport and mechanical properties is key in many applications, such as enhanced energy recovery and carbon geosequestration~\citep{Steefel_EPSL2005, Meakin_RG2009,Croizet2013, rohmer2016mechano, Xiong2016, carroll2016review}. 
%	
% The practical implications and scientific challenge of chemo-mechanical deformation, motivated intensive research, promoting our understanding of various aspects. 
%
% Much of the research has focused 
A large body of literature exists on rough fractures, due to their importance, e.g. dominance of fracture flow, and their quasi two-dimensional (2-D) nature, which simplifies experimental setup and imaging as well as numerical simulations~\citep[e.g.][]{Liu2006, detwiler2008experimental, elkhoury2013dissolution, ishibashi2013permeability, Ameli2014}. 
Relatively fewer works addressed porous media such as soils~\citep{cha2016hydro} and rocks~\citep[e.g.][]{Zheng2012, Croizet2013, Zheng2013, Emmanuel2015, buscarnera2016, nguyen2016, rohmer2016mechano,liu2017_2}, mostly focusing on the evolution of the mechanical properties.

% ######	research on chemomechanical , mostly focus on changes in mechanical peroperties . only few on permability ###### %%

The evolution of transport properties in a porous rock sample has been 
%The evolution of hydromechanical properties has been 
addressed by recent experiments, in which rock cores subjected to external stress were flushed by a reactive fluid~\citep{Vialle2011, Vanorio_Geophys2011, Grombacher2012, Vanorio2014, Vanorio2015, Clark2016}. These experiments show considerable reduction in mechanical stiffness, and inhibition of permeability enhancement, in spite of net removal of solid mass.
%
% In particular, samples with initially more compliant matrix (grain-supported or chalky carbonates) exhibited little permeability enhancement under stress, in spite of net dissolution and removal of solid mass~\citep{Vialle2011, Vanorio_Geophys2011, Grombacher2012, Vanorio2014, Vanorio2015, Clark2016}.
% This intriguing behavior cannot be explained by Existing models 
A model providing fundamental quantitative understanding of mechanisms governing the permeability evolution in these experiments---the objective of this paper---is lacking. 
Existing models of stressed rock dissolution are either at much larger scales (reservoir)~\citep{nguyen2016}, much smaller scales (describing changes in microstructure and porosity but not in permeability)~\citep{Emmanuel2015}, or consider other mechanisms such as pressure solution and re-precipitation~\citep{Zheng2012}.
%
%XXX--perhaps wait until identifying the gap---
%In this paper, we seek fundamental quantitative understanding of the mechanisms governing the permeability evolution in these experiments. 

% #####-- why challenge is ps. refer to recent review. then to ps num. models. ###### %%
 	%% ###### PORE SCALE MODELING ###### %%

A major challenge in our understanding of chemomechanical deformation is the dependency
%	The interplay between reactive transport and mechanical deformation strongly depends 
on pore-scale mechanisms and heterogeneity inherent to geologic media~\citep{RMG2015}. While the importance of pore-scale processes has long been recognized, rigorous studies are only recently made possible, owing to advancements in experimental and computational capabilities, as highlighted in a recent special volume of review papers~\citep{RMG2015}. The natural complexity of pore-scale processes associated with heterogeneity makes characterization~\citep[e.g.][]{Anovitz2015,Noiriel2015} as well as modeling and interpretation~~\citep[e.g.][]{Liu2015,Steefel2015,Molins2015,Yoon2015,Mehmani2015,smith2017} extremely challenging. This is especially true for cases where the coupling with mechanical deformation occurs~\citep[e.g.][]{Emmanuel2015,cha2016hydro,liu2017_2}. Despite of the advancements in experimental techniques
	% of controlling and characterizing flow and geochemistry and imaging the resulting pore space
	~\citep{Anovitz2015,Noiriel2015}, numerical simulations remain highly attractive~\citep[e.g.][]{Liu2015,liu2017_2,Steefel2015,Molins2015,Yoon2015,Mehmani2015,Xiong2016,smith2017}.

% ##### what we do here, in short ###### %%
	
%	In this paper, we seek fundamental understanding of the \textit{permeability evolution} during dissolution of stressed porous media. 
	%
	%This competition between pore opening by dissolution and pore closure associated with stiffness reduction, induced by the very same process--dissolution, and its impact on the permeability is a challenging problem in nonlinear physics with significant practical importance. 
	%
	In this paper, we present a novel pore-scale model, which, to the best of our knowledge, is the first to describe the permeability evolution caused by coupling between dissolution and compaction in a porous medium.  
	%
	%We focus on the conditions of the wormholing regime and at the transition to more uniform dissolution, attained when flow rate increased (lower $Da$). 
	Our simulations show that increasing stress inhibits the permeability evolution, requiring longer time or larger volumes injected to reach a certain permeability, in agreement with experiments~\citep{Vanorio_Geophys2011, Grombacher2012, Clark2016}. 
	We also expose the impact of the dissolution regimes ($Da$), that is how stress effect changes between wormholing and a more uniform dissolution. We find that at high Da, stress acts to reduce the transport and dissolution heterogeneity, promoting wormhole competition.

%[move to model description??]	Examples of complex chemo-mechanical deformation mechanisms include fracturing~\citep[e.g.][]{Royne2015, Clark2016}, sliding and detachment of mineral grains~\citep[e.g.][]{mangane2013, Clark2016}, pressure solution and welding of microporous zones~\citep{ishibashi2013permeability,Emmanuel2015, Clark2016}. 
%% 
%A simpler, yet fundamental process--stiffness reduction and subsequent compaction induced by dissolution--has recently been exposed by a series of core-flooding experiments~\citep{Vialle2011, Vanorio_Geophys2011, Grombacher2012, Vanorio2014, Vanorio2015, Clark2016}. The authors showed that application of stress led to a considerable reduction of stiffness and to inhibition of permeability enhancement, in spite of net removal of solid mass.
%% . In particular, samples with initially more compliant matrix (grain-supported or chalky carbonates) exhibited little permeability enhancement under stress, 
%%
%%XXX--perhaps wait until identifying the gap---
%In this paper, we seek fundamental quantitative understanding of the underlying mechanisms for the permeability evolution in these experiments. 

	% ==================================================
	% ==================================================
	% ==================================================
	\section{Pore-scale model} %ing of dissolution under stress}
	% ==================================================
	% ==================================================
	% ==================================================
	
	% ### WHAT PROCESS IS TAKEN INTO CONSIDERATION ### % 
	
	We simulate the injection of a reactive fluid into a porous medium held under a fixed confining stress. 
	We consider the gradual, continuous dissolution of the solid matrix and the consequent compaction, excluding mechanisms such as fracturing~\citep[e.g.][]{Royne2015, Clark2016}, detachment of mineral grains~\citep[e.g.][]{mangane2013, Clark2016}, and pressure solution and welding of microporous zoness~\citep[e.g.][]{ishibashi2013permeability, Emmanuel2015, Clark2016}. 	
	%
	% ### TYPE OF MODEL (PNM+BLOCK-SPRING) AND SIMPLIFIED GEOMETRY ### % 
	
\begin{figure}
		\includegraphics[width=.6\columnwidth]{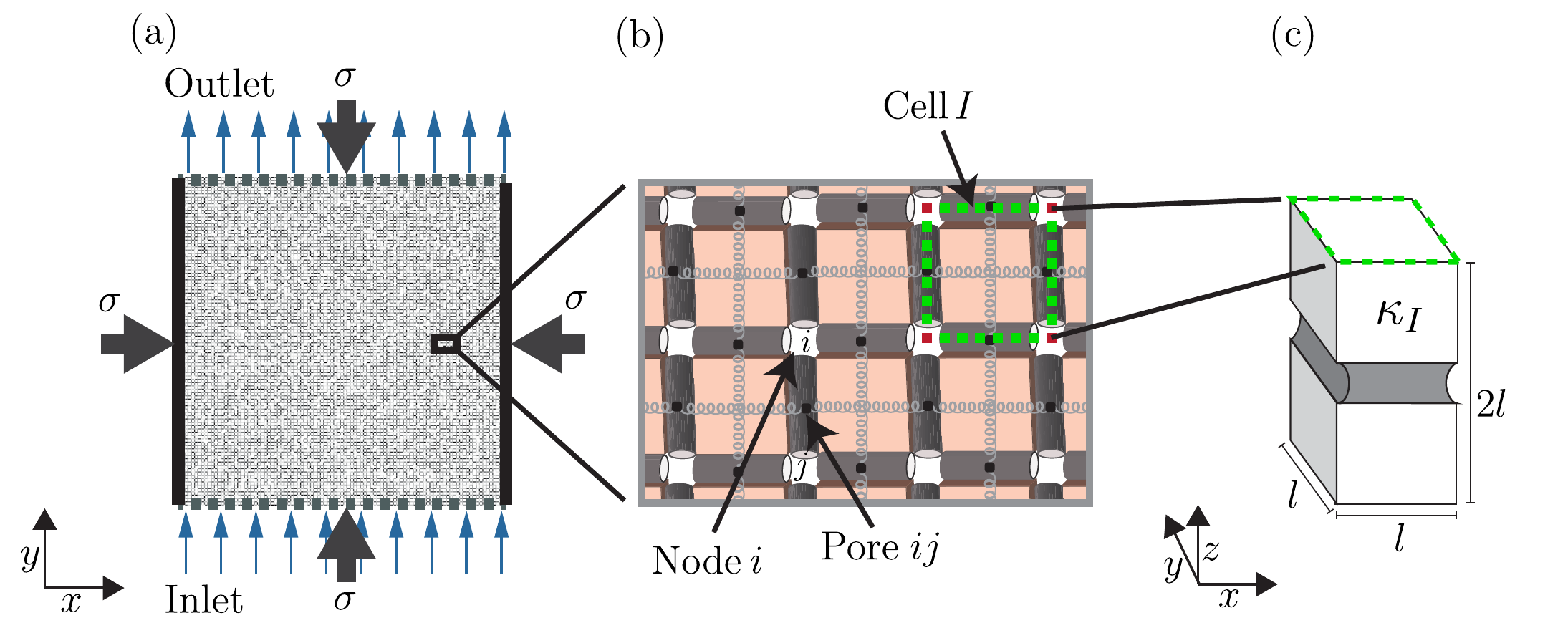}
		\centering
		\caption{Schematics of our pore-scale model of reactive transport in a stressed, deformable porous medium. We capture the coupling between chemical and mechanical deformation through a discrete representation, combining: (i)  pore network modeling of reactive transport, with (ii) interacting network of linear elastic blocks. 
		%, inducing : (i) solid removal (pore opening by chemical deformation) and (ii) weakening-induced compaction (pore closure by mechanical deformation). 
			%
			(a) We simulate the injection of a reactive fluid into a sample held under fixed isotropic stress, $\sigma$. 
			%, is being flushed by a reactive fluid, injected at a constant rate (in the $y$ direction). 
%
The sample is square, with two open faces (inlet and outlet, in the $y$ direction) and two impermeable ones (in $x$).
% that set the main flow direction to be in the $y$ direction. 
			%
			(b) The pore space is discretized into a 2-D regular network of cylindrical pores connected by volumeless nodes (or junctions), squeezed in between solid blocks. Heterogeneity is provided by variations in pore sizes.  
			(c) The basic model unit (``cell'') includes a solid block, with four parts of pores engraved in the solid. 
			%Each cell has the dimensions of $l${$\times$}$l${$\times$}$2l$ (\textcolor{green1}{We can remove this sentence and only keep the last one-- The out-of-plane.. .}).
			%
			The network properties, parametrized via the pore radius (assuming pores retain their cylindrical shape), are updated at each time step considering both the chemical and mechanical effects of dissolution.
} 
		\label{network}
	\end{figure}

	Seeking fundamental understanding rather than quantitative predictions for a specific type of medium, we use here an analog model with simplified geometry: a 2-D network of cylindrical channels between soluble solid blocks. 
	%, providing a tractable representation of the geochemical and mechanical behavior. 
%
	The coupling between dissolution and compaction at the pore-scale is captured by combining two discrete, overlapping representations: (i) pore network modeling of reactive transport and (ii) a network of interacting linear elastic porous solid blocks. 
	%
		%, providing a tractable representation of the geochemical and mechanical behavior. 
	Each discrete model unit (termed ``cell'' hereafter) includes one solid block, bounded by four ``half-channels''---the respective void space associated with that cell (Fig.~\ref{network}). 
	%
%	Dissolution of solid blocks alters the channel sizes by both chemical and mechanical deformation. 
	%induces two simultaneous, counteracting effects on the channels: increasing their sizes, and their compaction due to weakening.  
	%
	Our model can be considered as a realization of a granular sediment or a fractured rock, where each pore represents an interparticle void or a fracture, respectively.

	%%%%%%%%%%%%%%%%%%%%%%%%%%%%%

	Numerically, coupling is provided through staggering, executing sequentially at each time step the following steps: (1) evaluate the liquid pressures and fluxes,
	from mass conservation of liquid; (2) compute liquid composition, reaction rate
	and amount of dissolved mass, from conservation of solute mass; (3) compute stress distribution and pore deformation from force balance; and (4) update transport and mechanical properties (to be used in the subsequent time step) due to the combined effects of chemical pore enlargement by dissolution and mechanical compaction.
	%
%	the updated properties are then used in the next time step.
	%
	Our use of a quasi-static model via staggering is justified by the disparate time scales of the different processes: the time required to reach equilibrium of ion concentration is much longer than that for fluid pressures to equilibrate, which in turn is much slower than mechanical deformation and stress equilibration. % following a perturbation. 
	Therefore, the process hindering the rate of changes in the hydromechanical properties of the solid matrix is the dissolution, allowing us to treat the other processes as occurring instantaneously at every time step~\citep{Ameli2014}. That is, we approximate the continuous evolution of the sample properties by a discrete sequence of geochemical and mechanical equilibrium configurations.
	%
%	Furthermore, these processes occur over much shorter time scale than that required
%	to significantly alter the hydromechanical properties of the solid
%	matrix~\citep{Ameli2014}.\footnote{\textcolor{green1}{Check my proposition to this part. Perhaps this way it will be easier to understand that eventually the solid matrix deforms slow, and that chemical deformation rate mostly determines the rate of alteration of the hydromechanical properties.}} 
%	(\textcolor{green1}{"the time required to reach equilibrium of ion concentration is much longer than that for fluid pressures to equilibrate, which in turn occur over much shorter time scale than that required to significantly alter the hydromechanical properties of the solid matrix. Which deforms by slow dissolution and following quick stress equilibration and mechanical compaction, that occur over the shortest time scale".})
		%
	The different model components describing these processes are outlined below.

	\subsection{Reactive transport} \label{Reactive transport}
	
	We model the reactive transport and dissolution using a pore network model that closely follows the one in~\citet{Budek2012}; therefore, we only briefly describe it here, elaborating further on the coupling with mechanical compaction.

	\subsubsection{Fluid flow} 
	We consider a dilute liquid solution (e.g. CO$_2$-rich water), neglecting the effect of its pressure and solute concentration on its density.  
	%
	%Liquid fluxes are resolved from mass conservation.
	%
	%Exploiting the separation of timescales, we approximate the continuous evolution of sample properties by a discrete sequence of geochemical and mechaF
	For a given configuration (network geometry), we solve for the fluid fluxes by enforcing steady-state fluid mass conservation, namely the continuity equation, at each pore junction (node) $i$, 
	\begin{equation}
	{\sum_{j}} q_{ij}=0,
	\label{eq:Fluidbalance}
	\end{equation}
	where $q_{ij}$ is the volumetric flow rate through pore $ij$ ($q_{ij}>0$ indicates flow from node $i$ to $j$), and the summation is over all neighboring nodes, $j$ (Fig.~\ref{network}b). Assuming Stokes flow, the interpore fluxes, $q_{ij}$, are
	\begin{equation}
	q_{ij}=C_{ij} \Delta p_{ij}.
	\label{eq:Stokes}
	\end{equation}
	Here $C_{ij}= a_{ij} R_{ij}^{2}/(8\mu {l_{ij}})$ is the pore conductivity, $R_{ij}$ and $a_{ij} = \pi R_{ij}^{2}$ are the pore radius and cross-sectional area, and $\mu$ is the fluid viscosity. The term $\Delta p_{ij} = (p_i-p_j)$ is the pressure drop between the two nodes, which are $l_{ij}$ apart. 
	The linear system in Eq.~\ref{eq:Fluidbalance} provides the fluid pressure at the nodes. 
	Note that this formulation assumes that flow is driven by fluid pressure alone, i.e. a horizontal system; to generalize the formulation to include gravity effects, node pressures $p_i$ should be replaced by the total energy, e.g. in terms of hydraulic head.

	\subsubsection{Reagent transport and reaction rate} 
	
	The computation of dissolution rates is based on the following assumptions~\citep{Budek2012}: (1) steady-state liquid and solute transport; (2) solute transport in each pore is dominated by convection in the longitudinal (axial) direction, and by diffusion in the radial (transverse) direction; (3) reaction kinetics can be approximated via a linear law, $J_{r}=\lambda c_{w}$, where $\lambda$ is the surface reaction rate coefficient [LT$^{-1}$] and $c_{w}$ is the concentration at the pore surface; and (4) for 1-D radial diffusion, the diffusive flux can be evaluated from the difference between the surface concentration $c_{w}$ and cross-sectional averaged concentration (``mixing cup concentration'') $c$, by $J_{\rm{diff}}=D{Sh}(c-c_{w})/2R$. Here $D$ is the molecular diffusion coefficient, and $Sh$ is the Sherwood number.
	% [assigned here a fixed value of $Sh=4$~\citep{Budek2012}].

	The reaction rate, $J_{r}$, is expressed in terms of the \textit{pore-averaged} concentration $c_{ij}$, 
	\begin{equation} \label{eq:Jr}
	J_{r}=\lambda_{ij}^{\rm{eff}}c_{ij},
	\end{equation}
	where the effective reaction rate coefficient, $ \lambda_{ij}^{\rm{eff}} $, is computed assuming that it is set by radial diffusion between the pore center and the solid wall, $J_{\rm{diff}}=J_{r}$, providing 
	\begin{equation} \label{eq:eflamda}
	\lambda_{ij}^{\rm{eff}}=\frac{\lambda}{1+2\lambda R_{ij}/D{\rm{Sh}}}.
	\end{equation}
	The pore-averaged concentration, $c_{ij}$, is resolved from the 1-D steady-state analytical solution for solute mass conservation in each pore, providing an exponential concentration profile along the pore,
	\begin{equation}
	c_{ij}=c_{j}^{0}e^{ -A_{ij} \lambda_{ij}^{\rm{eff}} / q_{ji} },
	\label{eq:Solute} \end{equation}
	where $A_{ij} = 2\pi R_{ij}l_{ij}$ is the reactive surface area, and
	$c_{j}^{0}$ is the concentration at node $j$ evaluated at the previous time step (denoted by the superscript $0$). The node concentration is evaluated from the conservation of solute in each node,
	% Superscript $^0$ denotes a parameter value from the previous time step
	%
	\begin{equation}
	c_{i}=-\frac{\sum_{j({q_{ij}<0})}q_{ij}c_{ij}}{\sum_{j({q_{ij}>0})}q_{ij}}.
	\label{eq:Solute2} \end{equation}

	\subsection{Updating network properties due to chemomechanical deformation}
	
	We consider cylindrical pores which erode uniformly such that they retain their shape and length (here uniform, $l_{ij}=l$). This simplifies our computations, allowing us to parametrize the network properties such as the conductivity and reactive surface area by a single parameter, $R_{ij}$. 
	%
%	While $l_{ij}$ would also be altered by the dissolution, the effect of changes in the radius are far greater
	%
	%The properties are updated due to both solid removal (increasing the radius by $\Delta R_{ij}^{\rm{chem}}$), and compaction (decreasing it by $\Delta R_{ij}^{\rm{mech}}$), as described below. 
	%
	With staggering, the properties are updated at each time step considering the combined effects of (i) dissolution (increasing the radius by $\Delta R_{ij}^{\rm{chem}}$ relative to the previous time step) and (ii) compaction (decreasing it by $\Delta R_{ij}^{\rm{mech}}$)
, $R_{ij}=R_{ij}^0+\Delta R_{ij}$, where $\Delta R_{ij} = \Delta R_{ij}^{\rm{chem}} + \Delta R_{ij}^{\rm{mech}}$.
%, and $R_{ij}^0$ is the radius at the {previous} step.

	\subsubsection{Pore opening due to dissolution (chemical deformation)}\label{pore_opening}
	
	The computation of the direct effect of dissolution on pore geometry is based on conservation of solid mass. 
%	With the assumption that pores erode uniformly along their length, their radii change at every time step $\Delta t$ by
	The incremental change in radius at every time step $\Delta t$ is
	\begin{equation}
	\Delta R_{ij}^{\rm{chem}} = \frac{q_{ij}{\Delta t} c_{i}^{0}}{A_{ij}c_{\rm{sol}}\nu}\left(1-e^{-A_{ij} \lambda_{ij}^{\rm{eff}}/q_{ij}}\right)
	\label{eq:Vdiss}\end{equation}
	where $c_{\rm{sol}}$ is the concentration of soluble material at the surface, and $\nu$ accounts for the stoichiometry of the reaction~\citep{Budek2012}.
Our model accounts for two important aspects: the merger of adjacent pores following extensive dissolution, and that the amount of soluble solid is finite. 
The effect of merging is demonstrated in \citet{Budek2012} by comparing models with and without merging. However, these models do not limit the dissolution, considering an infinite amount of soluble solid.
	For simplicity, we use here the same threshold for merging and complete local dissolution. Thus, once all soluble solid associated with a cell has been consumed, (i) the dissolution reaction cease and the transport and mechanical properties of that cell remain fixed throughout the simulation; and (ii) each of the two pairs of pores associated with that cell (a pair in each direction, $x$ and $y$) are merged. 
	%
	
%\textcolor{blue}
Computationally, we model pore merger by an abrupt increase in the conductivity of the associated pores, while retaining the same discretization of nodes and pores. 
That is, the pair of merged pores are each assigned a conductivity $C_{ij}= C_{ij}^*$, which depends on the spatial extent of dissolution. 
%, that is, if a neighbor unit cell is also completely dissolved or not.
% (namely if an adjacent unit cell experienced complete dissolution or not).  
%
% Since each pore is associated with two solid blocks, the conductivity of the merged pores depends on the extent of the dissolution. 
%
When a cell experiences complete dissolution, while its adjacent neighbors did not, the conductivity of the pores in between them is set to 
$C_{ij}^* = a_{ij}^* {R_{ij}^*}^{2}/(8\mu {l_{ij}})$, where $R_{ij}^*=d_{ij}/2$, $d_{ij}=4a_{ij}^* / P_{ij}$ is the pore's hydraulic diameter, and $P_{ij}$ is its perimeter.
%
%To avoid counting twice 
%Keeping the same discretization, each pore in a merged pair is assigned with a conductivity. 
%
In evaluating $a_{ij}^*$ and $P_{ij}$ we associate the following regions to each pore: (1) half of a cylinder with radius $R_{ij}$, plus (2) half of the (completely dissolved) unit cell, common to both pores of that pair. 
%
%%% V9 pore still of where the two merged pores remain in place however with a much larger conductivity, we assign a large to avoid association of the same pore space twice, we evaluate $a_{ij}$ and $P_{ij}$ of each pore by assigning splitting the pore space equally--each pore is associated with half of the cell volume plus half of the volume of a cylinder with radius $R_{ij}=d_{ij}/2$.
%ROI: half of the dissolved unit cross-sectional area ($2l${$\times$}$l/2$), in addition to the area of the remaining half cylindrical channel (included in neighbour unit). When further dissolution takes place and the solid blocks from both sides of a pair of nodes is fully dissolved, we use conductivity between two parallel plates,  $C_{ij}=la_{ij}/(3\mu)$. 
	%
	To allow consideration of merging due to focused dissolution in a single pore (only from one side of the cell), we set the cell height (out-of-plane size) to be twice the lateral (in-plane) size, such that the cell dimensions are $l${$\times$}$l${$\times$}$2l$; with this, the radius of a single pore can approach $l$, and undergo merging (Fig.~\ref{network}c).
	Once two adjacent cells completely dissolve, the pore separating them is assigned with the conductivity of two parallel plates,
$C_{ij}^*=2 l^3/(3\mu)$.	
%	 $C_{ij}^*=l a_{ij}^*/(3\mu)$, with $a_{ij}^*=2 l^2$.
%
	
%	===== Below we briefly describe how we model merging; for further details, see ~\ref{pore_merging}. XXX =====\\
	%

	\subsubsection{Pore compaction (mechanical deformation)}  \label{mech}
	% #######################################################
	
	Pore compaction is evaluated by enforcing mechanical equilibrium (here, balance of forces). As before (for modeling reactive transport), we exploit the separation of time scales of compaction and dissolution to describe the continuous deformation as a sequence of static equilibrium configurations, each attained instantaneously.
	We assume that each cell is a linear elastic, homogeneous and isotropic solid, such that the stress and strain in each cell $I$ are related via Hooke's law,
	%
	%\begin{equation}\label{eq:block_strain}
	$\sigma_{I}^{\beta} = \kappa_{I} \epsilon^{\beta}_{I}$,
	%\end{equation}
	%
	where $\sigma^{\beta}_{I}$ and $\epsilon^{\beta}_{I}$ are the stress and strain in the $\beta$ direction ($x$ or $y$). The cell effective stiffness, $\kappa_{I}$, is evaluated using effective medium theory~\citep{MavkoHandbook2009},
	\begin{equation}\label{eq:ed_compressibility}
	\kappa_{I}=\left[\frac{1}{\kappa_{s}}+\frac{\phi_{I}}{\kappa_{\phi}}\right]^{-1},
	\end{equation}
	where $\phi_{I}$ is the cell's porosity---the ratio of pore to total volume---which changes with dissolution and compaction, $\kappa_{s}$ is the bulk modulus of the solid (a fixed material property), and $\kappa_{\phi}$ is the pore stiffness, which can be determined experimentally~\citep{Vanorio2014, Vanorio2015}. Our model considers compaction induced solely by changes in the local porosity, $\phi_{I}$, without accounting for the secondary effect of pore stiffness degradation~\citep{Vanorio2014, Vanorio2015}; thus, we use a fixed $\kappa_{\phi}$ value.

	We consider external confining stress on the sample applied via rigid walls that remain planar. While this imposes uniform displacement of all outermost (boundary) cells, the stress distribution on them becomes nonuniform due to the nonuniform stiffness resulting from the heterogeneous dissolution.
Finally, to compute the compaction, we invoke the following additional simplifying assumptions: (i) perfectly-drained conditions, without consideration of poroelastic effects, i.e. changes in pore pressure due to mechanical deformation; and (ii) deformations remain sufficiently small such that they can be decoupled in $x$ and $y$.
The above assumptions, together with superposition provided by linearity of the stress-strain relations, makes evaluation of compaction highly computationally efficient.

	%Decoupling of the deformations in $x$ and $y$ simplifies the computations of the boundary displacements which are required to achieve the desired macroscopic external stress. \footnote{\textcolor{green1}{We already mentioned that we decouple the deformations in $x$ and $y$ in previous paragraph, perhaps we should make change here.}} 
	
	%, such that each row or column of in the network bearer load according to its effective stiffness and further local erosion and weakening of the pore-units changes the stress distribution.  
	
	The computational procedure used to establish mechanical equilibrium and associated compaction is as follows. At each time step we compute: (1) the stiffness of each cell, $\kappa_{I}$, using Eq.~\ref{eq:ed_compressibility}; (2) the effective sample stiffness in each direction, $\kappa^{\beta}_{\textrm{eff}}$; (3) the boundary displacements (macroscopic strains), $\epsilon^{\beta}_\textrm{eff}= \sigma / \kappa^{\beta}_{\textrm{eff}}$, given the macroscopic external stress applied on the boundary, $\sigma$; 
	% (4) evaluate the stress on each row or column of cells, $\sigma_s$, noting that the decoupling in $x$ and $y$ and the fact that the boundaries remain planar provides $\sigma_s = \kappa_s \epsilon_s$, with $\epsilon_s = \epsilon^{\beta}_\textrm{eff}$, $\kappa_s$ being the effective stiffness of row or column $s$; (5) use force balance to infer the stress distribution on individual cells, 
%	
(4) the stress on cell $I$, $\sigma^{\beta}_{I}$, noting that due to the decoupling in $x$ and $y$, force balance reduces to the equality of the stress on all cells in a given row of cells $s$, $\sigma^{\beta}_{I} = \sigma_s$. The stress in row $s$ is computed from $\sigma_s = \kappa_s \epsilon_s$ using the effective stiffness of each row, $\kappa_s$, where the planarity of the boundaries implies uniform strain of all rows in a specific direction, $\epsilon_s = \epsilon^{\beta}_\textrm{eff}$;
(5) the radial strain of each pore, $\epsilon_{ij}= \sigma_{ij} / \kappa_{{\phi}}$, assuming the radial stress on a pore is equal to the one acting on the cell, $\sigma_{ij} = \sigma^{\beta}_{I}$; and 
(6) the decrement in pore radius from geometrical arguments,
	\begin{equation}\label{eq:dR_mech2}
	\Delta R_{ij}^{\rm{mech}}=R_{ij}^{\rm{'}}\sqrt{1 -\epsilon_{ij}} - R_{ij}^0
	\end{equation}
where $R_{ij}^{\rm{'}}$ is the radius considering dissolution alone (excluding compaction) since the \textit{beginning of the simulation}, and $R_{ij}^0$ is the radius considering both dissolution and compaction; both $R_{ij}^{\rm{'}}$ and $R_{ij}^0$ are the known values from the \textit{previous time step}.
Note that the radius of a pore can experience a decompaction within a time step, $\Delta R_{ij}^{\rm{mech}}<0$, if stress redistribution leads to local release of stress in that area.
% (prior to the current incremental changes $\Delta R_{ij}^{\rm{mech}}$ and $\Delta R_{ij}^{\rm{chem}}$).

	% ######################################################
	\subsection{Initial and boundary conditions}
	% ######################################################
	\label{int and bc}
	
	Simulations begins at geochemical and mechanical equilibrium, with zero reagent concentration, $c(t=0)=0$.
	%, and unit cells sum of forces equal to zero, $\Sigma F_{I}=0$ (with sample under confining stress, $\sigma$). 
	We inject a liquid at fixed concentration $c= c_{in}$ and flow rate $Q$. We fix the rate throughout the dissolution---as permeability changes---by adjusting the pressure at the inlet nodes at every time step (keeping zero pressure at the outlet nodes). 
 Injection occurs from the inlet face towards the outlet face, as the two perpendicular faces are impermeable (see Fig.~\ref{network}a).
At the outlet pores, we enforce free-flow boundary conditions, computing their concentration from solute mass conservation (Eqs.~\ref{eq:Solute}--\ref{eq:Solute2}).
	Mechanically, we apply a constant confining stress on the boundaries, $\sigma$, keeping the boundaries planar; that is, we enforce uniform displacement of the boundary cells, which therefore experience non-uniform forces (depending on the stiffness). 
	%(stress is aplied through planar rigid walls).
	
	%To summarize, we use the following initial and boundary conditions for the reactive transport (pore-network) model: $c(i,t=0)=0$, $c(i_{\rm{in}},t>0)=c_{in}$, $\Sigma q(i_{\rm{in}},t>0)=Q_{in}$, $q({ij}_{\rm{imp}},t>0)=0$, where subscript in and imp refer to a boundary node (pore junction) located at the inlet or at the impermeable sides, respectively. For the mechanical (block-spring) model the initial and boundary conditions are: $\Sigma F(I,t=0)=0$ and $\sigma (\Gamma,t \geq 0)$, where $\Gamma$ indicates boundary.

	\subsection{Parameter values used in the simulations}\label{parameter_values}
	
	In all simulations, we used a domain of 100$\times$100 nodes. We chose a square sample with an aspect ratio of one to minimize the interference of the lateral boundaries with wormhole development~\citep{cohen2008pore}. 
	%Square sample rather than rectangular one is chosen to reduce boundary effects which can affect wormhole competition \citep{cohen2008pore,kalia2009}. 
%	
	We enforce an initial heterogeneity by assigning the pores with volumes drawn from a uniform distribution, $V_{ij}\in[1-\alpha, 1+\alpha]\bar{V}$, where $\bar{V}$ is the average volume and $\alpha$ is a measure of the heterogeneity, here $\alpha=0.8$.
	The initial overall sample porosity is $\phi=0.1$. 
	 The amount of insoluble solid is a fixed, spatially-uniform \textit{fraction} of the total cell volume, $\rho=0.2$. 
	 Due to the initial heterogeneity in pore sizes, solid volume and porosity vary among cells.

	% 0.8\footnote{VERIFY THIS IS TRUE!}\footnote{\textcolor{green1}{ It's not, the soluble solid volume in each unit is not uniform, because the pore sizes are not uniform. In the units there is insoluble volume fraction of 0.2, pores of variable size and the rest is soluble solid.}}. 
	
	The reactive transport conditions can be described by two dimensionless groups \citep{Budek2012}: (i) ${Da}$, the ratio between the reaction rate and the mean flow rate in pores; and (ii) ${G}$, the extent to which the dissolution rate is hindered by diffusive transport of reactant from the pore to the wall,
	\begin{equation}
	{{Da}} = \frac{2\pi \bar{R} l \lambda /\bar{q}_{i_{\rm{in}}}}{1+\lambda 2 \bar{R} /D{\rm{Sh}}}
	\label{eq:DA} \end{equation} 
	and
	\begin{equation}
	{G} =\lambda 2 \bar{R} /D{\rm{Sh}},
	\label{eq:G}\end{equation} 
	where $\bar{q}_{i_{\rm{in}}}$ and $\bar{R}$ are the average flow rate at the inlet and the initial average radius, respectively. 
	We use a fixed value of ${Sh}=4$~\citep{Budek2012}. 
%In accordance, the dimensionless time is defined as $\hat{t}=\lambda c_{in}t/(\bar{R}\nu c_{\rm{sol}})$.
	%
	We focus on two regimes: (a) heterogeneous dissolution or wormholing with $Da=1$, and (b) a more uniform dissolution, $Da=0.01$. We control $Da$ via the total flow rate, $Q_{\rm{in}}$, while fixing $G=1$. We note that compaction due to the application of stress changes slightly the values of $Da$ and $G$ (by up to $\sim$5\% and $\sim$20\%, respectively). 
	
	%. Additionally, we note that following initial loading the values of $Da$ and $G$ slightly change (by $\textasciitilde 5\%$ by $\textasciitilde 20\%$, respectively), i.e. aside from the compacting radius, the parameters constituting $Da$ and $G$ (Eq.~\ref{eq:DA} and \ref{eq:G}) remains constant between simulations of different stress conditions and uniform initial $Da$. 
	
    To simulate carbonate dissolution we set $c_{in}=0.001$ M; this corresponds to pH=3, for which the system is sufficiently far from geochemical equilibrium to justify the approximation of first order reaction kinetics. We use a molecular diffusion coefficient of $D$=3{$\cdot$}$10^{-9}$ m$^2$/s and a rate coefficient of $\lambda$=5{$\cdot$}$10^{-4}$ m/s \citep{peng2015kinetics,alkattan1998}. 
    % This simple description of the reaction kinetics captures the main geochemical effect, while avoiding the complexity when all details of calcite dissolution are considered. 
%    
	% This simplistic description of the complex calcite dissolution kinetics is adopted to gain fundamental insights rather than provide accurate predictions for a specific set of medium and fluids. 
	%
	Mechanically, we apply a fixed stress, $\sigma$, ranging between 0 and 60MPa. We set stiffness of $\kappa_{s}=22$ GPa and $\kappa_{\phi}=0.01 \kappa_{s}$, characteristic of soft sediments \citep{Vanorio2014, Vanorio2015}. 
	
	We keep the numerical error associated with staggering and time discretization small by using a sufficiently small time step $\Delta t$, 20 and 0.5 s for $Da$=1 and 0.01, respectively. This value was chosen by trial and error, running a set of simulations with decreasing time step until the pore volume to breakthrough ($PV_{BT}$) stabilized.
	%, both for conditions of no stress and under stress. 
%
	Breakthrough is defined here as a ten-fold increase in the permeability, $K/K_{0}$=10, where $K$ and $K_{0}$ are the current and initial permeability.
	
%	\footnote{HOW DO YOU DEFINE AN ERROR HERE? E.G., YOU COMPARE SEVERAL SIMULATIONS WITH DIFFERENT TIME STEPS AND CLAIM THE ONE WITH SMALL ENOUGH STEP SUCH THAT RESULTS DO NOT CHANGE IS ACCURATE? AN EXPLANATION IS DUE HERE..\textcolor{green1}{At the original version it is explained that the error grows linearily with time step, so the zero error associated with time step length can be extrapolated. But that requires some lengthy explanations}. \textcolor{red}{So?? it is still unexplained in the current version, hence cannot be published.. Explain to me in brief simple language and I'll find the way to put it here. Anyhow, I suspect ``error grows linearily with time step'' is neither accurate nor well-explained in previous versions, as again--to talk about error you need first to define the accurate answer, which we don't have here analytically..}} 
%	
	%time step length analysis we made indicates that under no stress the error with respect to pore volumes injected until breakthrough, $PV_{BT}$, increases linearly with time step length and updating frequency. For conditions of $Da=1$ and $0.01$ and $G=1$ the appropriate dimensionless time step length, with error smaller than $2\%$ in $PV_{BT}$ under all stress conditions is $\Delta\hat{t}=3E-5$ and $8E-7$, respectively. 
	
% ==================================================
% ==================================================
% ==================================================
\section{Results}
% ==================================================
% ==================================================
% ==================================================

\subsection{Permeability evolution}

% ==================================================
% RESULTS (1): PERMEABILITY EVOLUTION inhibited by stress
% ==================================================

\begin{figure}
	\includegraphics[width=.7\columnwidth]{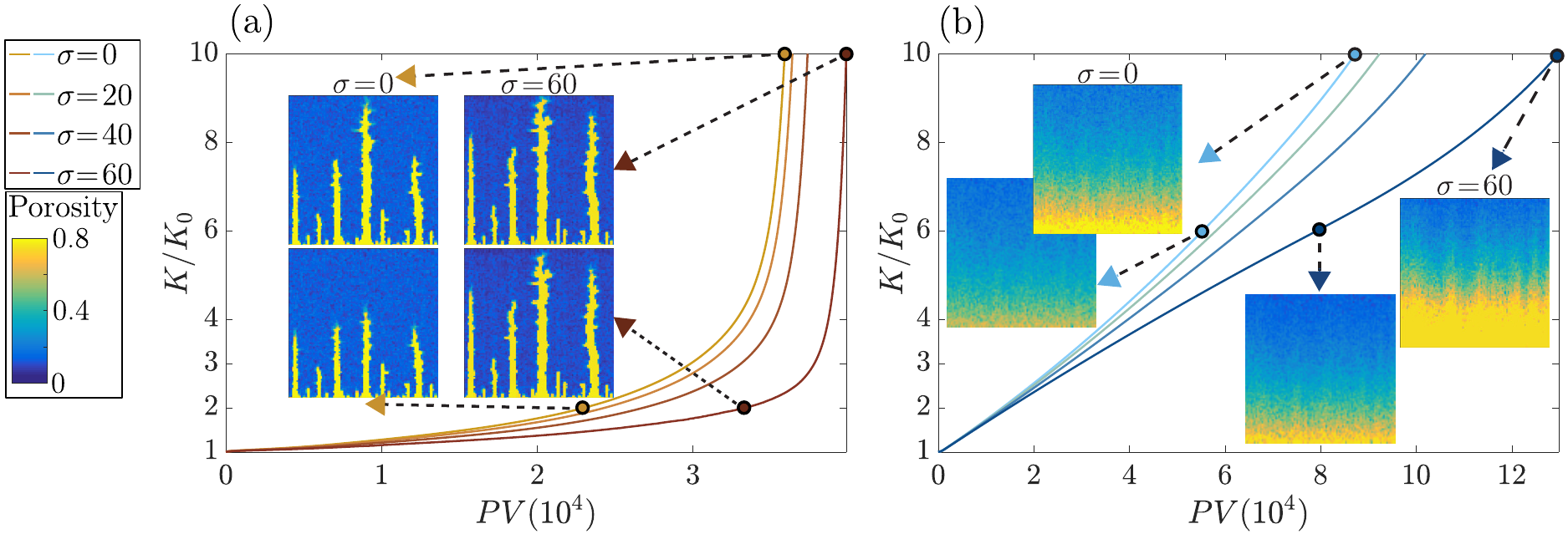} % ELSEVIER
	\centering
	\caption{Our simulations demonstrate the inhibition of permeability enhancement by stress. 
	% effect of stress at different dissolution regimes: $Da$=1 (a) and 0.01 (b). 
	Increasing stress, $\sigma$, lowers the dissolution efficiency, such that a larger volume of fluid, $PV$ (or, equivalently, longer time), is required to reach a given permeability, $K/K_0$. 
		Consequently, the dissolution front progresses further downstream at a given $K/K_0$ value (see insets).
		At early times (low $K/K_0$), stress effect is more noticeable at $Da$=1, delaying the sharp rise in permeability. For example, the difference in $PV$ required to double the permeability ($K/K_0$=2) under $\sigma$=60 MPa vs. $\sigma$=0, is about five times larger for $Da$=1 ($\sim$45\%, panel a) than for $Da$=0.01 ($\sim$10\%, b).
		This trend is reversed later on: at breakthrough ($K/K_0$=10), the increase in $PV_{BT}$ is about five times larger at $Da$=0.01 than at $Da$=1.
		The curves represent ensemble averages from multiple realizations (see text). %For each stress, the data represents an ensemble average of 600 realizations (different samples); the large number of realizations was required due to the large sensitivity of the results to the specific details of the sample geometry, with a standard deviation ranging between $\sim$4000--5000 $PV$ (increasing with stress).
	}
	\label{PV-perm} 
\end{figure}

Our simulations demonstrate the intricate interplay between chemical and mechanical deformation, capturing the inhibition of permeability enhancement associated with stress-induced compaction, observed experimentally~\citep{ Grombacher2012, Vanorio2014,Vanorio2015,Clark2016}. 
%, a mechanism that becomes more and more dominant as stress increases. 
%
Increasing the external stress results in longer time, or equivalently larger amounts of injected fluid, $PV$, required to reach a certain permeability (Fig.~\ref{PV-perm}). 

Permeability and its inhibition by compaction strongly depends on the spatial extent of the dissolution, e.g. on the regime ($Da$), and changes with time. 
We quantify this through the increase in $PV$ required to reach a certain $K/K_0$ under high stress of $\sigma$=60 MPa relative to $\sigma$=0, $\Delta PV = [PV_{\sigma=60}-PV_{\sigma=0}]/PV_{\sigma=0}$. 
At early times (low $K/K_0$), the impact of stress is more noticeable for wormholing ($Da$=1); for instance, at $K/K_0$=2 the increment {$\Delta PV$} is $\sim$45\% for $Da$=1 vs. only $\sim$10\% for $Da$=0.01.
This trend however flips later on: at breakthrough, {$\Delta PV$} is five times larger for $Da$=0.01, with {$\Delta PV$}$\approx$50\% (Fig.~\ref{PV-perm}). 
For both regimes, the reduction in dissolution efficiency (increase in $PV$) associated with stress causes the dissolution front to progress further downstream (e.g. compare insets in Fig.~\ref{PV-perm} at a given $K/K_0$).
Regardless of stress, permeability enhancement remains much more efficient at high $Da$, consuming about a third of the reagent volume at breakthrough.

%% While this result is expected, our simulations point to
%We expose a rather complex, unintuitive interplay with the flow velocity ($Da$), affecting the governing mechanisms of dissolution under stress. The effect of stress is initially more notable for high $Da$, however at later times (higher permeability) becomes more apparent at low $Da$. 

%The dissolution behavior in general, and the permeability evolution in particular, strongly depend on the dissolution regime. Under no stress (neglecting mechanical deformation), the behavior changes from relatively uniform dissolution, and a relatively constant rate of permeability enhancement ($\sim$linear with time or injected pore volumes, $PV$) for fast injection, to more heterogeneous, preferential dissolution at slower injection rates~\citep{Daccord_nature1987, Bekri_CES1995, Hoefner_AIChE1988, Szymczak_JGR2009,upadhyay2015}.

%% (in terms of time$\hat{t}$, 0.27-0.3 and 0.0065-0.01 for $Da$=1 and 0.01, respectively\footnote{ROI what are these numbers? range for different stresses? please explain}\footnote{\textcolor{green1}{It's dimensionless time at breakthrough and yes, the range is for different stresses.}}). 

% ==================================================
% ==================================================

The curves in Fig.~\ref{PV-perm} are obtained by ensemble averaging over multiple realizations, namely different sample geometry with similar \textit{statistical} properties. 
As wormholing instability is more sensitive to the initial conditions than uniform dissolution---showing greater variability between realizations, more realizations were required to obtain a statistically-representative result: 200 for $Da$=1, vs. only 15 for $Da$=0.01. 
This provided a standard error in $PV_{BT}$ smaller than 1\% (relative to the mean), and a standard deviation of $\sim$10\% and $\sim$1\% for $Da$=1 and $Da$=0.01, respectively.

\subsection{Stress distribution}
% ==================================================
% RESULTS (3): EXPLAINING MECHANICAL EFFECT ON PERMEABILITY VIA  EVOLUTION OF STRESS DISTRIBUTION 
% ==================================================

%%%% STRAIN %%%%

\begin{figure}[h]
	\centering
		\includegraphics[width=.6\columnwidth]{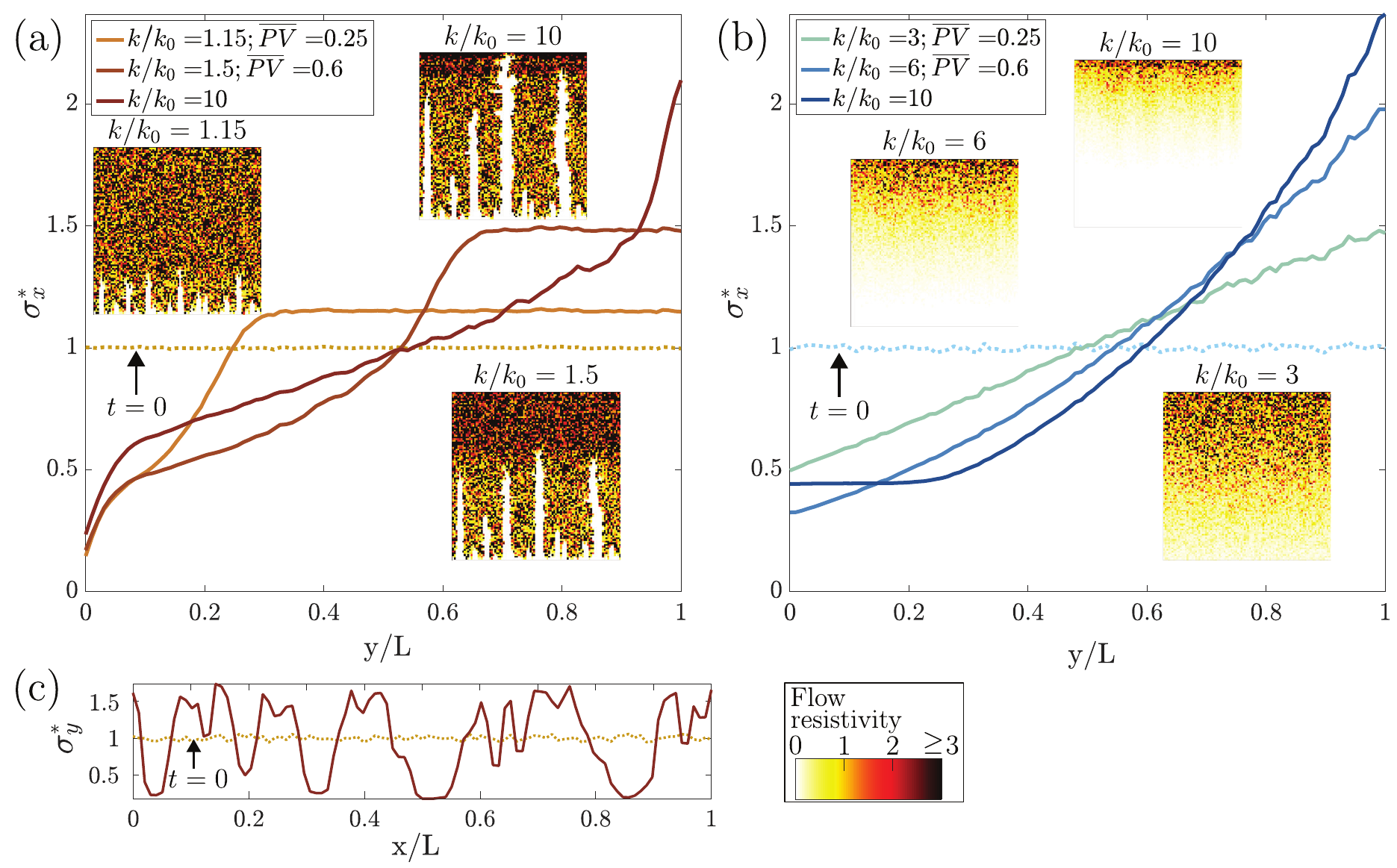} % ELSEVIER
	\caption{We explain the permeability evolution through the spatiotemporal distribution of stress. Intense dissolution upstream increases its compliance, hence the load carried by the stiffer downstream region. 
		Since the latter is also less conductive (insets), its compaction provides a strong bottleneck effect that constrains permeability. 
		%Insets show longitudinal flow resistance, $1/C_{ij}$, normalized by the initial ($t$=0) average resistance.
		%
	    (a) For $Da$=1, the sharp dissolution contrast between the upstream and downstream regions amplifies the bottleneck effect at early times (low $K/K_0$).
		%
%Wormhole propagation progressively relieves stress upstream, increasing the stress and hence compaction on the downstream region. 
%
		Once the dissolution (i.e. main wormhole) reaches the outlet, the effect of this mechanism becomes less noticeable.
		(b) In contrast, for $Da$=0.01, the incomplete reaction allows the dissolution to propagate throughout the sample, leading to a much smaller conductivity contrast. 
		% This gradual softening leads to a gradual increase in stress with high stress concentrated in a \textit{narrow} region near the outlet (Fig.~\ref{Stress_dist}b).
		%, early on (at relatively low $K/K_0$; see Fig.~\ref{Stress_dist}a). 
		%
		Dissolution at the outlet induces both pore enlargement and its partial counteraction by compaction, with a steady, continuous increase in the effect of stress on permeability. Near breakthrough, the impact of stress at $Da$=0.01 overtakes that at $Da$=1 (Fig.~\ref{PV-perm}).
		Panels a--b show the local stress (normalized by the macroscopic stress, $\sigma$=60 MPa) in the \textit{transverse} direction, $\sigma^{*}_{x}$, responsible for compacting pores along the main flow direction. %, which control permeability. 
		(c) Stress in the \textit{longitudinal} direction, $\sigma^{*}_{y}$ (shown at $t$=0 and breakthrough), despite its high heterogeneity (attaining a maximum at the stiffer regions between the wormholes), has a much smaller effect on permeability. 
		%
%		The local stresses are normalized by the externally applied macroscopic stress, here $\sigma$=60 MPa. 
		%
		%Insets show resistivity normalized by the averaged initial pore resistivity (see text). %
		Each curve represent an ensemble average at a given $K/K_0$ values; the corresponding volumes (normalized by $PV_{BT}$), $\overline{PV}$, are also provided in the legend.
}
	\label{Stress_dist} 
\end{figure}

We explain the complex interplay between dissolution and compaction through the dynamic evolution of stiffness and stress distribution. 
While matrix dissolution in a region increases its compliance, the very same process has another, opposite effect: it decreases the load that region bears, since stiffer regions carry more of the load. 
The competition between these two mechanisms controls the compaction, affecting permeability.
In particular, the permeability is governed by the least conductive region; here, this is the downstream region, which---being furthest away from the inlet---experiences the least dissolution. 
In both regimes, intensive dissolution upstream significantly increases the load downstream (Fig.~\ref{Stress_dist}a--b). 
Even minute compaction of the less conductive downstream region (see flow resistance in insets of Fig.~\ref{Stress_dist}) intensifies the bottleneck effect, limiting the permeability.  
	Insets in Fig.~\ref{Stress_dist} show longitudinal resistance, $1/C_{ij}^y$, normalized by the initial ($t$=0) average resistance.

%which is furthest away from the inlet, where the reagent is introduced, since the reagent is introduced from the opposite side (inlet) and the outlet experiences the least amount of dissolution.
% (being furthest away from where the reagent is introduced), 

%
%We note that the dynamic evolution of the stiffness and stress distribution is highly nonlinear, and complex: while dissolution of a region decreases its stiffness, it also decreases the load it carries. The competition of these two effects determines the amount of compaction, which, depending on its location, determines the overall effect on the permeability.

%
%Since this region experiences less dissolution, it is also less conductive (acting as a bottleneck that controls permeability); therefore, stress application intensifies the bottleneck effect, reducing permeability.
%
The effect of stress depends on its spatial distribution, which in turn depends on the evolving dissolution pattern. 
For $Da$=1, dissolution is focused within the wormholes, creating a distinct dissolution front. 
%behind this front dissolution is massive and hence it bears little load, whereas 
%
Ahead of this front remains a \textit{wide} undissolved and hence stiffer zone (extending from the most advanced wormhole tip to the outlet), upon which most of the load is distributed (Fig.~\ref{Stress_dist}a). 
%
%\footnote{\textcolor{green1}{We does not mention that as the wormholing progresses this stiffer region shrink but stress concentration there governes, curbing the permeability enhancement expected by the wormholes progression. I think we have enough extra words to explain that. } THIS APPEARS IN THE NEXT SENTENCE..\textcolor{green1}{Most of it does not appear.}}\footnote{\textcolor{green1}{We should indicate that despite the stress concentration, wormholes release some of the load and continue to propagate.}}
%
The elevated stress on this region, which hardly experiences dissolution until breakthrough, significantly retards the permeability enhancement (insets of Fig.~\ref{Stress_dist}a). This decreases the transport heterogeneity, promoting wormhole competition (cf. Section~\ref{sect_wormhole_comp}).  
As the wormholes propagate, the stress is progressively relieved from the newly dissolved region, and increases downstream. 
Once the outlet starts eroding, the focused dissolution characteristic of wormholing becomes dominant, diminishing the bottleneck effect and providing a sharp permeability increase. 

%\footnote{\textcolor{green1}{Perhaps: "Once the outlet starts eroding by the focused dissolution characteristic of wormholing, it diminishes the bottleneck effect and breakthrough quickly reached.".}}
 %(Fig.~\ref{PV-perm}).
%

% ==================================================
% ==================================================

The behavior changes at $Da$=0.01: the rapid injection and incomplete reaction allows some reagent to propagate further downstream, resulting in a much more uniform dissolution and mechanical weakening throughout the sample. 
%This spatially-gradual softening leads to gradually increasing stress with high stress concentrated in a \textit{narrow} region near the outlet (Fig.~\ref{Stress_dist}b).
%, early on (at relatively low $K/K_0$; see Fig.~\ref{Stress_dist}a). 
%
The continuous dissolution at the outlet region, experiencing both pore enlargement as well as its partial counteraction by compaction, leads to a steady, relatively-linear increase in the effect of stress on permeability (increasing the divergence among the curves in Fig.~\ref{PV-perm}b). The effect of stress, which is initially larger at $Da$=1, becomes more noticeable at $Da$=0.01 as breakthrough is approached.

%Consequently, the effect of stress at breakthrough (e.g. in $PV_{BT}$) becomes much less noticeable relative to that in $Da$=0.01. 

%%
%The local stress curves of Fig.~\ref{Stress_dist}a-b, averaged for constant permeability enhancement values, $K/K_0$, where $\bar{PV}$ is the respective average $PV$.
%%
%The insets show resistivity of pores along the main flow direction ($y$), normalized by the initial average pore resistivity ($1/C_{ij}$).  

In our 1-D flow settings, pores parallel to the flow direction ($y$) mostly control the permeability. Therefore, we plot in Fig.~\ref{Stress_dist}a--b the local stress in the transverse direction, $\sigma^{*}_x$.
%, that are associated with their compaction. 
%
The highly heterogeneous distribution of stress in the longitudinal direction, $\sigma^{*}_{y}$, where the stress is concentrated at the stiffer regions between the wormholes~\citep{cha2016hydro} (Fig.~\ref{Stress_dist}c), has only a minor effect on permeability.

\subsection{Wormhole competition}\label{sect_wormhole_comp}

\begin{figure}[h]
	\centering
		\includegraphics[width=.5\columnwidth]{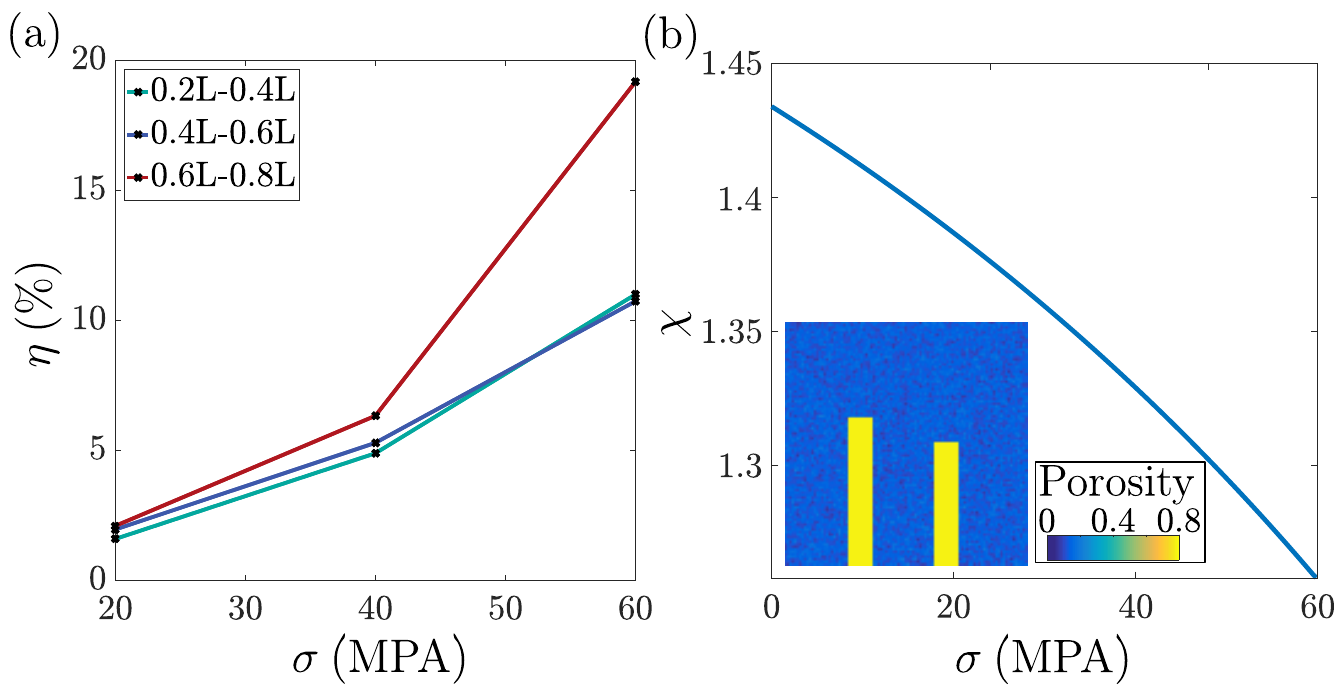} % ELSEVIER
	\caption{Stress increases wormhole competition, thus the number of wormholes. 
		(a) At breakthrough, the relative increment of number of wormholes (vs. $\sigma$=0), $\eta$, increases with stress throughout the sample, in particular downstream (at $y$=0.6-0.8$L$). 
		$\eta$ represents an ensemble average measured in three bands of width of 0.2$L$.
		This behavior is due to a decrease in transport heterogeneity, where reagent is being distributed more evenly throughout the sample. It inhibits the propagation of the most developed wormhole, thus delaying the sharp permeability increase associated with wormholing at breakthrough. 
		%
		%To evaluate $\eta$, we count the number of contiguous regions which have been completely dissolved at each longitudinal distance ($y$$\in$$[0,1]L$), and compute the difference (compared to $\sigma$=0) averaged along three bands of width of 0.2$L$. Every value is an ensemble average over 200 realizations. 
		%
		(b) We exemplify the underlying mechanism via a simple model system consisting of two straight wormholes (of length $0.6L$ and $0.5L$, see inset). We show that the flow rate through the more advanced one, normalized by that in the shorter one, $\chi$, decreases with stress, due to induced relative compaction ahead of the main wormhole.   
		This demonstrates that stress decreases the transport heterogeneity, promoting wormhole competition which is responsible for the slower breakthrough.}
	\label{wrmh_number} 
\end{figure}

A further quantitative insight into the impact of stress on the complex dynamics at high $Da$ can be provided from analyzing the pattern evolution. Stress increases the competition among wormholes, hence the number and length of secondary wormholes on the expense of the dominant wormhole (e.g. insets in Fig.~\ref{PV-perm}a). 
Fig.~\ref{wrmh_number}a shows the relative increment in the number of wormholes at different confining stresses (compared to $\sigma=0$), $\eta$, measured in three regions of width of 0.2$L$ (centred at $y$=0.3, 0.5 and 0.7$L$). A wormhole is defined here as a contiguous region which has been completely dissolved. 
% The value of $\eta$ is obtained from ensemble averaging over 200 realizations. 

% ==================================================
% ==================================================

Heterogeneous dissolution due to heterogeneous transport, where the most advanced wormhole increasingly draws more fluid and thus propagates at the expense of less developed neighbors, is the mechanism responsible for the sharp increase in permeability (Fig.~\ref{PV-perm}a). 
Application of stress reduces the transport heterogeneity, 
%such that the reagent is more uniformly distributed, 
%
slowing down the propagation of the main wormhole, thus breakthrough.
We demonstrate this quantitatively via a simple system consisting of two straight wormholes (Fig.~\ref{wrmh_number}b, inset), measuring the ratio of flow rates through the long and short wormholes, $\chi$. 
Increasing stress is shown to divert less of the flow into the main wormhole (lowering $\chi$), indicating a reduction in the degree of preferential fluid flow---and hence reagent transport (Fig.~\ref{wrmh_number}b).

Interestingly, stress was found to have an opposite effect---increasing transport heterogeneity---in a different system: \textit{non-reactive} transport in a rough fracture~\citep{Kang2016}. The authors show that stress increases the number of solid-solid contacts, forcing the solute to pass through fewer, hence more preferential pathways. 
In our system, stress \textit{reduces} the conductivity contrast among the wormholes, hence the dominance of the main wormhole. That is, stress promotes dissolution of smaller wormholes, reducing transport heterogeneity and dissolution efficiency.

\section{Summary and conclusions}

We study the permeability evolution in a stressed, deformable porous sample undergoing dissolution. 
We present a novel pore-scale model describing the coupling between (a) pore opening (chemical deformation) and (b) mechanical weakening and pore compaction. Our simulations point to a complex, unintuitive effect of stress. As the upstream region undergoes substantial dissolution, higher load is being carried by the stiffer downstream region. Since this region is also less conductive, even its small compaction has a significant bottleneck effect, such that larger injected volume (or equivalently, longer time) is required to reach a certain permeability. 

The manner by which the permeability enhancement is curbed by stress depends on the dissolution regime ($Da$).
At high injection rates (low $Da$), the relatively uniform dissolution (including downstream) leads to a steady increase in the effect of stress, caused by continuous dissolution at the outlet and its partial counteraction by compaction. 
In the wormholing regime (high $Da$), the undissolved region ahead of the dissolution front has an acute bottleneck effect. This acts to reduce transport heterogeneity, by decreasing the contrast between wormhole conductivity which promotes wormhole competition. At early times (low permeability), this mechanism strongly suppresses the permeability enhancement, much more than at low $Da$. Once the main wormhole approaches the outlet, this mechanism diminishes and focused reagent transport and dissolution leads to a sharp permeability rise. Consequently, close to breakthrough the impact of stress becomes larger at low $Da$.  
%
%stress reduces transport heterogeneity, hence increasing wormhole competition and delaying the sharp permeability rise typical of wormholing. The inhibiting effect of stress, is therefore larger at high $Da$ at early times, a trend which flips towards breakthrough.
%
Our work improves the understanding of how the hydromechanical properties change with reactive transport, a process with important implications in natural engineered systems ranging from diagenesis and weathering of rocks, to well stimulation and carbon geosequetration.

%% &&&&&&&&&&&&&&&&&&&&&&&&&&&&&&&&&&&&&&&&&&&&&&&&&&&&&&&&&6
%% &&&&&&&&&&&&&&&&&&&&&&&&&&&&&&&&&&&&&&&&&&&&&&&&&&&&&&&&&6
%%%%%%%%%%% APPENDIX %%%%%%%%%
%\renewcommand{\thesection}{Appendix \Alph{section}}
%\setcounter{section}{0}
%\renewcommand{\thefigure}{A\arabic{figure}}
%\renewcommand{\theequation}{A\arabic{equation}}
%\setcounter{figure}{0}
%\setcounter{equation}{0}
%
%%\appendix
%
%
%% ==================================================
%\section{Modeling pore merging}
%%\section*{(i) Derivation of the Liquid Volume-Meniscus Curvature Relationship}
%\label{pore_merging}
%
%Text here. With the commands I inserted here, Sections, Figures and Eq. will have special enumeration, automatically: Fig.~\ref{fig:merging}, Eq.(\ref{temp})
%
%	\begin{equation}
%	A=B
%	\label{temp} 
%	\end{equation}
%	
%% ==================================================
%\begin{figure}[h]
%	\centering
%%	\includegraphics[width=0.4\columnwidth]{SpheriCap.pdf}
%	\caption{Pore merging... 
%}
%	\label{fig:merging}
%\end{figure}
%% ==================================================
%% &&&&&&&&&&&&&&&&&&&&&&&&&&&&&&&&&&&&&&&&&&&&&&&&&&&&&&&&&6
%% &&&&&&&&&&&&&&&&&&&&&&&&&&&&&&&&&&&&&&&&&&&&&&&&&&&&&&&&&6

% &&&&&&&&&&&&&&&&&&&&&&&&&&&&&&&&&&&&&&&&&&&&&&&&&&&&&&&&&6
\subsection*{Acknowledgments}
%% \acknowledgments % AGU
%\begin{acknowledgments} % ELSEVIER
We gratefully acknowledge financial support by the United States-Israel Binational Science
Foundation (BSF-2012140), Israeli Science Foundation (ISF-867/13), and Israel Ministry of Agriculture
and Rural Development (821-0137-13). We also thank T. Vanorio and A. Nur for valuable discussions. 
%\end{acknowledgments}

%% %%%%% ARXIV
%\bibliographystyle{model2-names.bst}\biboptions{authoryear}
%\bibliography{dissolution_under_confining_stress}
%

\end{document}